\documentclass[12pt]{article}
\usepackage{epsfig}
\usepackage{hhline}

\begin{document}

\title{Extended self-similarity of the
small-scale cosmic microwave background anisotropy}

\author{A. Bershadskii$^{1,2}$ and K.R. Sreenivasan$^2$}

\maketitle
\begin{center}
1.   {\it ICAR, P.O. Box 31155, Jerusalem 91000, Israel}

2.   {\it ICTP, Strada Costiera 11, I-34100 Trieste, Italy}
\end{center}

\begin{abstract}
The Extended Self-Similarity (ESS) of cosmic microwave background
(CMB) radiation has been studied using recent data obtained by the
space-craft based Wilkinson Microwave Anisotropy Probe. Using the
ESS and the high angular scale resolution (arcminutes) of the data
it is shown that the CMB temperature space {\it increments}
exhibit considerable and systematic declination from Gaussianity
for high order moments at the small angular scales. Moreover, the
CMB space increment ESS exponents have remarkably close values to
the ESS exponents observed in turbulence (in magnetohydrodynamic
turbulence).

\end{abstract}



\newpage

Interaction of the primordial magnetic filed with the
baryon-photon fluid is widely discussed as a possible origin of
the space anisotropies of the cosmic microwave background (CMB)
radiation \cite{beo}-\cite{b}. Recent analysis of the COBE data
\cite{b} showed that for very large cosmological scales (larger
than $10^o$) the Anderson localization makes it impossible for the
electromagnetic fields to propagate. However, for the angular
scales less than $4^o$ the electromagnetic fields can be actively
involved into macroscopic dynamics (including turbulent motion of
the baryon-photon fluid \cite{beo}, \cite{beck}-\cite{bs}). The
release of the first results from the Wilkinson Microwave
Anisotropy Probe (WMAP) can be very helpful for this problem. The
high angular resolution (arcminutes) and space-craft origin of the
WMAP data make the WMAP temperature maps very attractive. \\

 In present Letter we will use a map made from the WMAP data
 \cite{tag}. The observations were made at frequencies 23, 33, 41,
 61, and 94 GHz. The entire map consists of 3145728 pixels, in thermodynamic
 millikelvins.  The map was cleaned from foreground contamination
 (due to point sources, galactic dust, free-free radiation, synchrotron
 radiation and so forth). Since some pixels are much noisier than
 others and there are noise correlations between pixels, Wiener filtered map
 is more useful than the raw one. Wiener filtering suppresses the noisiest
modes in a map and shows the signal that is statistically
significant.  \\

The main tool for extracting cosmological information from the CMB
maps is the angular power spectra. However, such spectra use only
a fraction of the information on hand. An additional tool is the
probability density function (PDF) of the CMB temperature {\it
fluctuations}, or deviations from a mean value. The WMAP
temperature fluctuations have been confirmed to be Gaussian. Now,
taking advantage of the good angular resolution of this map, we
can study statistical properties of angular {\it increments} of
CMB temperature for different values of angular separation. The
increments are defined as
$$
\Delta T_r = (T({\bf R}+{\bf r}) - T({\bf R}))   \eqno{(1)}
$$
where ${\bf r}$ is dimensionless vector connecting two pixels of
the map separated by a distance $r$, and the structure functions
of order $p$ as $\langle|\Delta T_r|^p \rangle$ where
$\langle.\rangle$ means a statistical average over the map.

It is well known that for complex stochastic (turbulent) systems
the temperature fluctuations $\delta T = T-\langle T \rangle$
themselves can exhibit rather good Gaussianity, while the space
increments (1) are significantly non-Gaussian. So-called Extended
Self-Similarity is an useful tool to check non-Gaussian properties
of the data. The ESS is defined as the generalized scaling (see,
for instance, Ref. \cite{ben} for application of the ESS to
different turbulent processes):
$$
\langle|\Delta T_r|^p \rangle \sim \langle|\Delta T_r|^3
\rangle^{\zeta_p} \eqno{(2)}
$$
For any Gaussian process the exponents $\zeta_p$ obey to simple
equation
$$
\zeta_p = \frac{p}{3}   \eqno{(3)}
$$
Therefore, systematic deviation from the simple law (3) can be
interpreted as deviation from Gaussianity. An additional
remarkable property of the ESS is that the ESS holds rather well
even in situations when the ordinary scaling does not exit, or
cannot be detected due to small scaling range (that takes place in
our case) \cite{ben}.

Figure 1 shows logarithm of moments of different orders
$\langle|\Delta T_r|^p \rangle$ against logarithm of
$\langle|\Delta T_r|^3 \rangle$  for the cleaned and
Wiener-filtered WMAP data. The straight lines (the best fit) are
drawn to indicate the ESS (2).

\begin{figure}[ht]
\epsfig{file=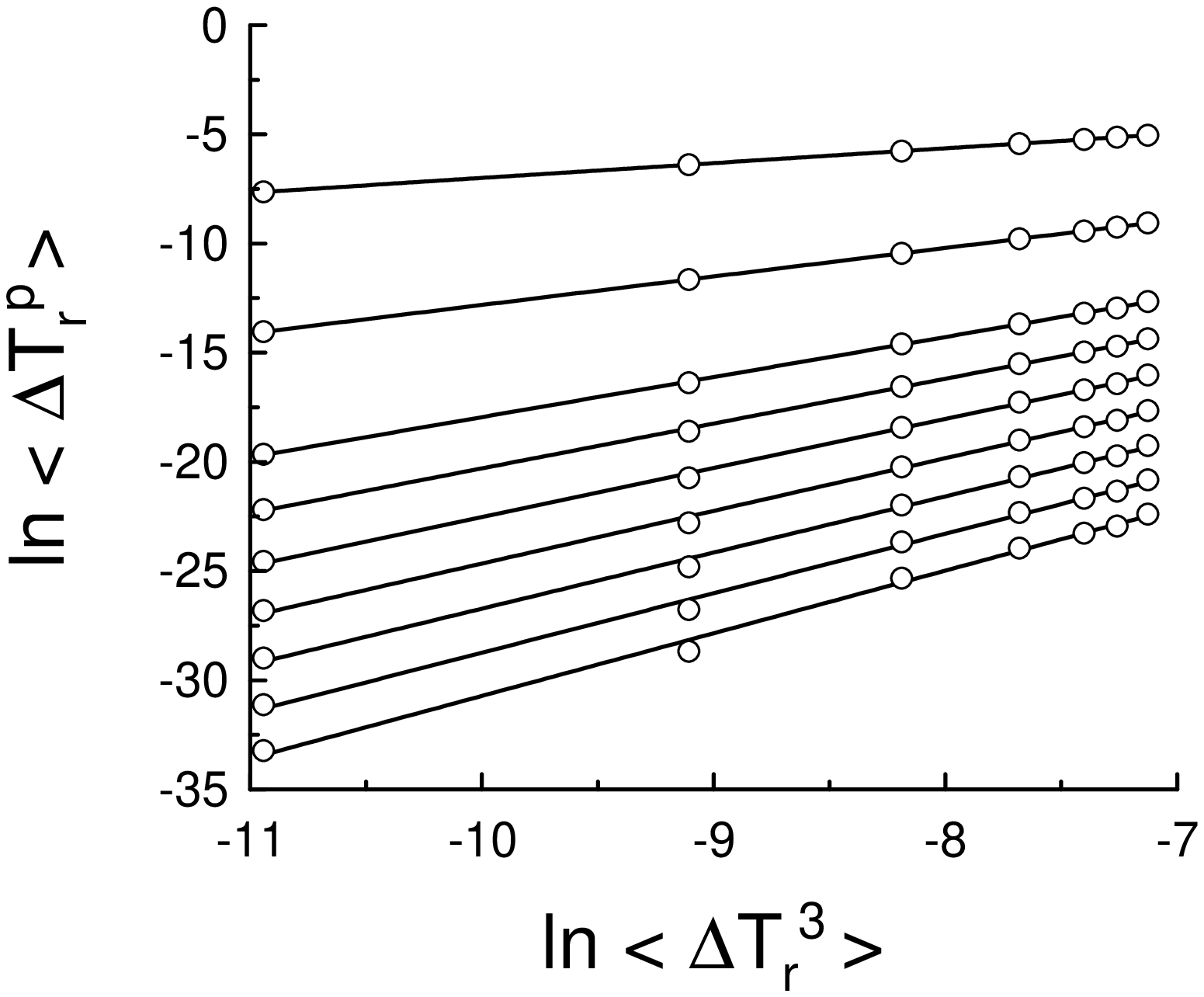,width=4.5in} \caption{\footnotesize
Logarithm of moments of different orders $\langle|\Delta T_r|^p
\rangle$ against logarithm of $\langle|\Delta T_r|^3 \rangle$  for
the cleaned and Wiener-filtered WMAP data. The straight lines (the
best fit) are drawn to indicate the scaling (2).}
\end{figure}

Figure 2 shows the exponents $\zeta_p$ (open circles) extracted
from figure 1 as slopes of the straight lines. We also show
(dashed straight line) in figure 2 dependence of $\zeta_p$ on $p$
for the Gaussian distributions (3).

\begin{figure}[ht]
\epsfig{file=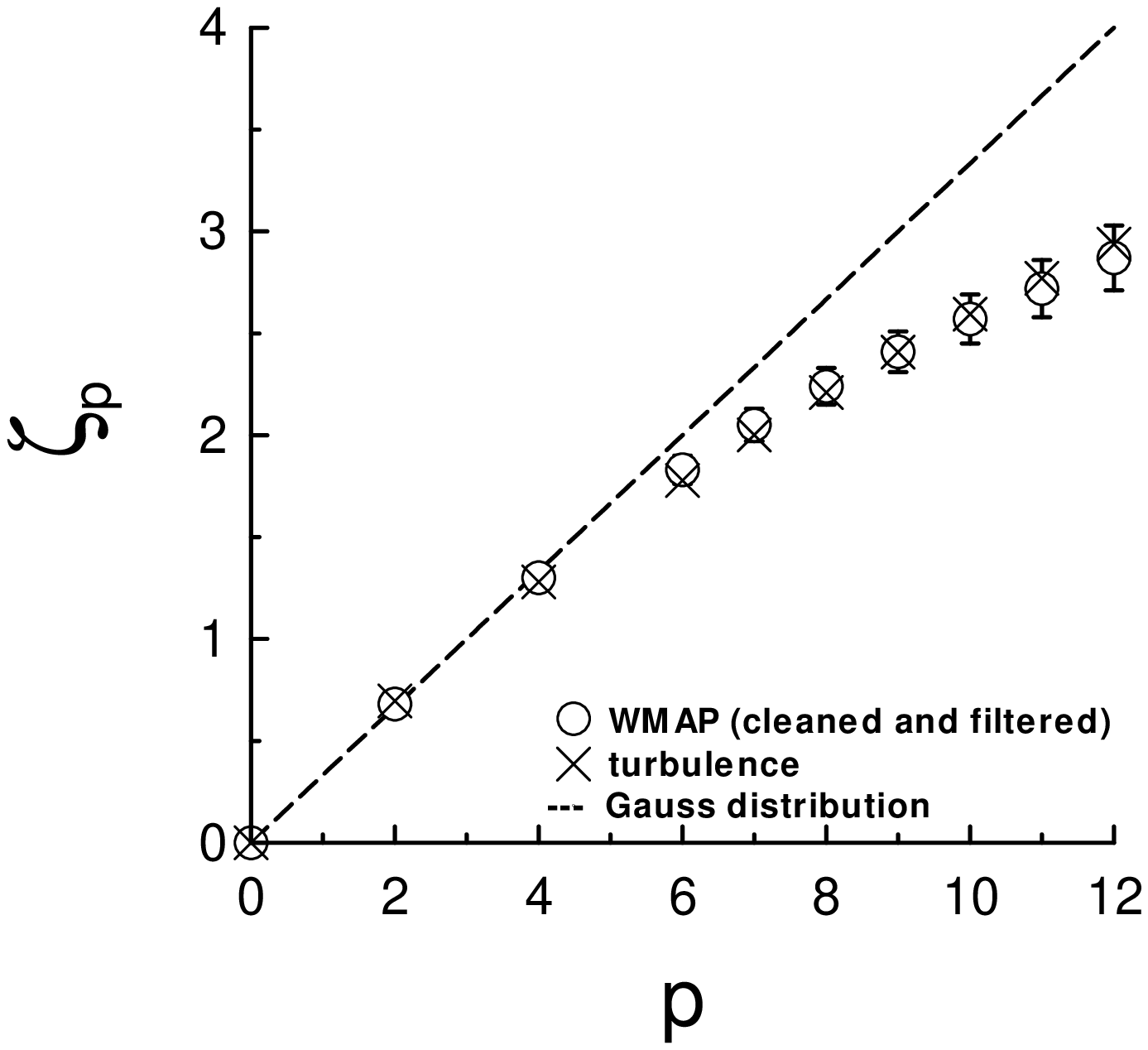,width=4.5in} \caption{\footnotesize The
exponents $\zeta_p$ (open circles) extracted from figure 1 as
slopes of the straight lines. The dashed straight line corresponds
to the Gauss distribution ($\zeta_p=p/3$), while the crosses
correspond to the turbulence ESS \cite{ben}.}
\end{figure}

One can see that starting from $p \simeq 6$ the data decline
systematically from the Gaussian straight line and follow, in this
declination, quite close to the ESS exponents observed for
turbulence \cite{ben} (indicated in the figure by the crosses). \\

Since the quoted "turbulent" values of the exponent $\zeta_p$ are
practically the same both for classic Kolmogorov fluid turbulence
and for the Alfven-wave dominated magnetohydrodynamic (MHD)
turbulence \cite{ben} it is impossible, using the presented data,
decide - what type of turbulence (Kolmogorov's or MHD) could be
considered as the CMB modulation origin in this case. According to
the up to date knowledge of the processes in the cosmic
baryon-photon fluid the last (MHD) type of turbulence seems to be
more plausible \cite{beo}-\cite{dky}. It is know now that the
Alfven waves with small enough wavelength, which can oscillate
appreciably before recombination, become overdamped (the photon
mean-free-path becomes large enough for dissipative effects to
overcome the oscillations). In this case, the longest wavelength
which suffers appreciable damping by photon viscosity, has a scale
$k^{-1}\sim V_A L_S^C$, where $L_S^C$ is the usual comoving Silk
(or photon viscosity) scale and $V_A$ is the Alfven speed. Since
the Alfven speed is $V_A\sim 3.8\times 10^{-4}B_{pr}<<1$ ($B_{pr}$
is the present-day magnetic field in units $10^{-9}$ Gauss), only
comoving wavelengths smaller than $10^{23}B_{pr}$ $cm$ suffer
appreciable damping. Therefore, it seems plausible that the
arcminute WMAP data could be modulated by the Alfven-wave
dominated turbulence, that results in the evident systematic
declination of the small-scale WMAP temperature space increments
from the Gaussianity
(figure 2).\\

The authors are grateful to C.H. Gibson for discussions, to M.
Tegmark group for providing the data, and to the Referee for
suggestion.


\newpage


\begin{thebibliography}{99}
\bibitem{beo} A. Brandenburg, K. Enqvist and P. Olesen,
Phys. Rev. D, {\bf 54}, 1291 (1996).
\bibitem{sb} K. Subramanian and J.D. Barrow, Phys. Rev.
Lett., {\bf 81}, 3575 (1998).
\bibitem{jko} K. Jedamzik, V. Katalinic, and A.V. Olinto,
Phys. Rev. D, {\bf 57}, 3264 (1998).
\bibitem{sb3} K. Subramanian and J.D. Barrow.
Phys. Rev. D, {\bf 58}, 083502 (1998).
\bibitem{dky} R. Durrer, T. Kahniashvili, and A. Yates,
Phys. Rev. D, {\bf 58}, 123004 (1998).
\bibitem{b} A. Bershadskii, Phys. Lett. B, {\bf 559} (2003) 107.
\bibitem{beck} R. Beck, A. Bradebburg, D. Moss, A. Shukurov, D.
Sokoloff, Annu. Rev. Astron. Astrophys., {\bf 34}, 153 (1996).
\bibitem{dolgov} A.D. Dolgov, D. Grasso and A. Nicolis,
Phys. Rev. D, {\bf 66}, 103505 (2002).
\bibitem{kos} A. Kosowsky, A. Mack, and T. Kahniashvili, Phys. Rev. D,
{\bf 66}, 024030 (2002).
\bibitem{bs} A. Bershadskii and K.R. Sreenivasan, Phys. Lett. A,
{\bf 299}, 149 (2002).
\bibitem{tag} M. Tegmark, A. de Oliveira-Costa, A. Hamilton,
astro-ph/0302496.
\bibitem{ben} R. Benzi, L. Biferale, S. Ciliberto, M.V. Struglia,
R. Tripiccione, Physica D, {\bf 96}, 162 (1996).


\end{thebibliography}
\end{document}